\documentclass[aps,prapplied,reprint,groupedaddress]{revtex4-2}

\usepackage{graphicx}
\usepackage{dcolumn}
\usepackage{bm}
\usepackage{float}

\begin{document}


\title{Bias-free reconfigurable magnonic phase shifter based on a spin-current controlled ferromagnetic resonator}


\author{Zikang Zhang}
\author{Shuang Liu}
\author{Tianlong Wen}
\author{Dainan Zhang}
\author{Lichuan Jin}
\author{Yulong Liao}
\author{Xiaoli Tang}
\author{Zhiyong Zhong}
\email[]{zzy@uestc.edu.cn}
\affiliation{State Key Laboratory of Electronic Thin Films and Integrated Devices, University of Electronic Science and
Technology of China, Chengdu 610054, China}


\date{\today}

\begin{abstract}
Controllable phase modulation plays a pivotal role in the researches of magnonic logic gates. Here we propose a reconfigurable spin-current controlled magnonic phase shifter based on a ferromagnetic resonator. The proposed phase shifter requires no magnetic bias field during operation. The device is directly configured over the waveguide while keeping the original structure of the waveguide unaffected. Numerical micromagnetic simulations show that the phase shifter could yield either a $\pi$-phase or no shift depending on the magnetization status of the resonator, which can be controlled by a current pulse. Moreover, the phase-shifting operation could be affected by spin current. At different input current density, the device could be either used as a dynamic controlled phase shifter or a spin-wave valve. Finally, a XNOR magnonic logic gate is demonstrated using the proposed phase shifter. Our work can be a beneficial step to enhance the functionality and compatibility of the magnonic logic circuits.
\end{abstract}


\maketitle

\section{INTRODUCTION}
Further miniaturization and integration of complementary metal oxide semiconductor (CMOS) circuitry are limited by the fundamental physical limitation and the Joule heating dissipation \cite{1,2}. This circumstance stimulates a global interest in search for novel alternative technologies of CMOS \cite{3,4}. Among these alternative technologies, using spin waves (or magnons) as information carriers, which could be used to fulfill wave-based information processing \cite{54}, is one of the most promising technologies \cite{5,6,7,8,9,10,11}. The typical operational wavelength of spin waves is several orders of magnitude shorter than that of the electromagnetic waves at the same frequency  \cite{8,10,12,13}, which allows for a better dimensional scaling of magnonic devices in nano-meter scale. Besides, due to no particle transfer during the transmission of spin waves, magnonic devices have extremely low energy consumptions \cite{14}, which was estimated to be at the aJ level by Intel’s benchmarking of beyond-CMOS devices \cite{15,16}. Furthermore, with encoding the information into the amplitude and phase of spin waves, magnonic devices could even open the way to non-Boolean computing \cite{17,18,19,20,21}, reversible logic \cite{22,23,24} and artificial neural networks \cite{25,26,54}.
\par
Recent studies have highlighted the promising application of magnonic devices in logic circuits \cite{14,27,28,29,30,31,32}. Historically, the most representative prototype of magnonic logic gates was based on the principle of the Mach-Zenhder-type interferometer (MZI) \cite{17}. In this prototype logic gate, the spin waves (SWs) propagate along the two branches of the MZI and interfere at the output terminal. By modulating the phases of the SWs in the two branches, the output signal was interfered either constructively or destructively depending on the relative phase difference of the two incident SWs. Here the magnonic phase shifter is responsible for the phase modulation and the interference output, which is critical for magnonic logic gates. As a result, an efficient reconfigurable magnonic phase shifter is essential for the efficient implementation of the interferometer-based magnonic logic gates. Various phase-shifting mechanisms have been proposed in recent years. The main approaches are based on the micromagnetic structure such as the domain walls \cite{33,34,35}, magnetic defects \cite{36,37}, magnonic crystals \cite{38} and others \cite{47,48}. These micromagnetic structure based magnonic phase shifters generally serve as passive elements to provide constant phase shifts without extra energy input, namely, they are "static". These static magnonic phase shifters are not adaptable for the dynamic occasions. However, in the construction of magnonic logic gates, we do need the dynamic phase shifter, which could be controlled in real time by an external signal. The reported approaches to dynamic phase shifter includes that applying magnetic fields \cite{27,39}, electric currents \cite{40}, spin-polarized currents \cite{41,42,43} or electric fields \cite{32,44,45,46} to achieve the externally initiated phase modulation. The main requirements for these dynamic phase shifters include simple control mode, low power consumption and scalable structure \cite{14}. From this point of view, most dynamic magnonic phase shifters now have drawbacks such as the performance degradation without bias field, tremendous input energy and the narrow application scope.
\par
In this paper, we propose a bias-free magnonic phase shifter based on a ferromagnetic resonator. The resonator is dynamically modulated by a relatively small magnitude of spin current. The proposed device could be used as a static or dynamic magnonic phase shifter depending on the operation mode.
\begin{figure*}
\includegraphics{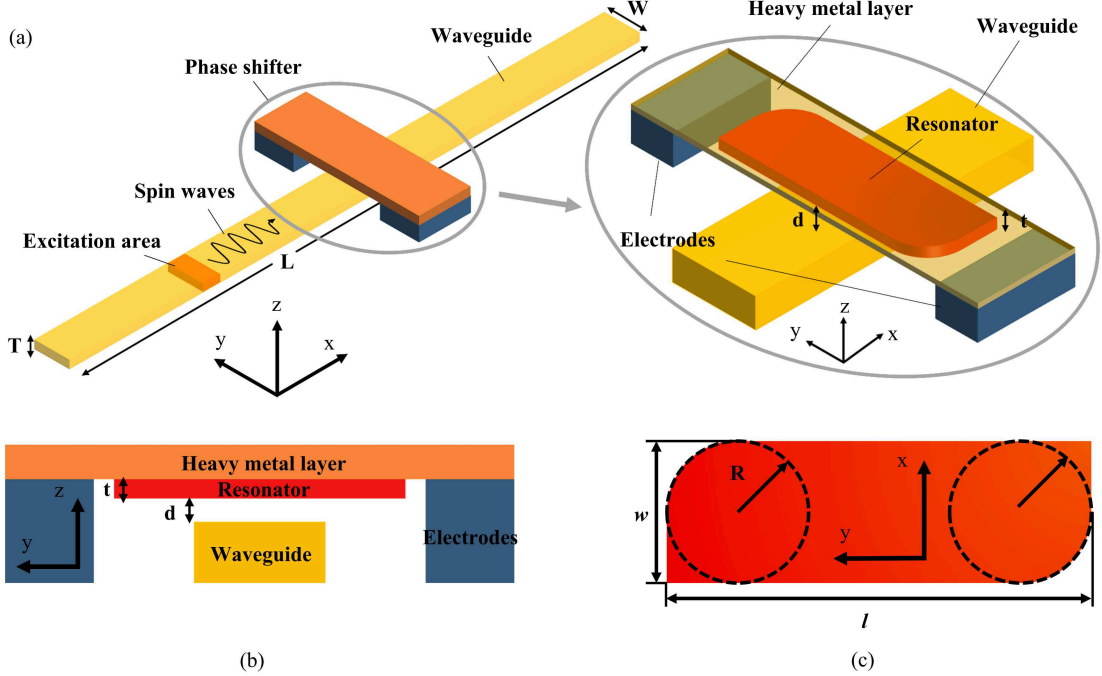}
\caption{\label{fig1} Device structure. (a) Schematic illustration of the micromagnetic model and perspective image (zoom in) of the magnonic phase shifter. The magnonic phase shifter includes a heavy metal layer with dimensions of 100 nm $\times$ 300 nm $\times$ 10 nm, a resonator with the dimensions of 150 nm ($l$) $\times$ 50 nm ($w$) $\times$ 2 nm ($t$) and electrodes. An excitation field with 200 A/m amplitude generates spin wave along the x-axis. The length, width and thickness of stripe waveguide are L = 2000 nm, W = 50 nm and T = 4 nm, respectively. (b) Side view of the waveguide and phase shifter. The resonator is formed beneath the heavy metal layer, and the spacing between resonator and waveguide is d = 2 nm. (c) Detailed parameters of the resonator. The two diagonal corners of the resonator are transformed into an arc with radius R = 25 nm.}
\end{figure*}
\par
The remainder of this paper is organized as follows. In Sec. II, the device structure and the simulation methods are presented. In Sec. III, we first give an overview of the phase-shifting mechanism, and we then perform numerical simulations to demonstrate a controllable static phase shifter mode and a dynamic phase shifter mode. Finally, we highlight the implementation of universal Boolean XNOR logic gate and the applicable conditions of the proposed magnonic phase shifter. Sec. IV summarizes the key results and advantages of this work.

\section{METHODS}
Fig.1 shows the scheme of the proposed magnonic phase shifter. The phase shifter is consisted of a narrow strip waveguide with dimensions of 2 $\mu$m (L) $\times$ 50 nm (W) $\times$ 4 nm (T), two blocks of electrode, a heavy metal layer (e.g. Pt, Td, W, etc.) with dimensions of 100 nm $\times$ 300 nm $\times$ 10 nm and a resonator. Both the waveguide and the resonator are made of Permalloy (Py), which is suitable for nano-patterning \cite{14}. The length, width and thickness of the resonator are $l$ = 150 nm, $w$ = 50 nm and $t$ = 2 nm, respectively. In order to destroy the axial symmetry, two diagonal corners of the resonator are designed to be an arc with radius R = 25 nm. The resonator is placed above the center of the waveguide with spacing between them of d = 2 nm, which is achieved by the supporting of electrodes. We performed the simulation using Object Oriented Micromagnetic Framework (OOMMF) that numerically solves the Landau-Lifshitz-Gilbert (LLG) equation \cite{49}. The cell size is set as 5 $\times$ 5 $\times$ 2 nm$^3$. The following material parameters of Py are used: saturation magnetization $M_s$ = 8 $\times$ 10$^5$ A/m, exchange stiffness $A$ = 1.3 $\times$ 10$^{-11}$ J/m and anisotropy constant $K$ = 0 J/m$^3$. The Gilbert damping factor $\alpha$ is chosen to be 0.5 for obtaining equilibrium state, 0.005 for dynamic simulations and 0.05 at the ends of the waveguide to prevent the reflection of spin waves. The ground state is obtained by relaxing the system under no bias field for the waveguide and uniformly magnetized for the resonator in the positive x and y axis directions respectively. We choose a sinusoidal magnetic field along the y axis with 200 A/m amplitude, 10.50 GHz frequency as the excitation signal. The excitation area is 50 $\times$ 50 nm$^2$ at 500 nm away from the left end of the waveguide.

\begin{figure*}
\includegraphics{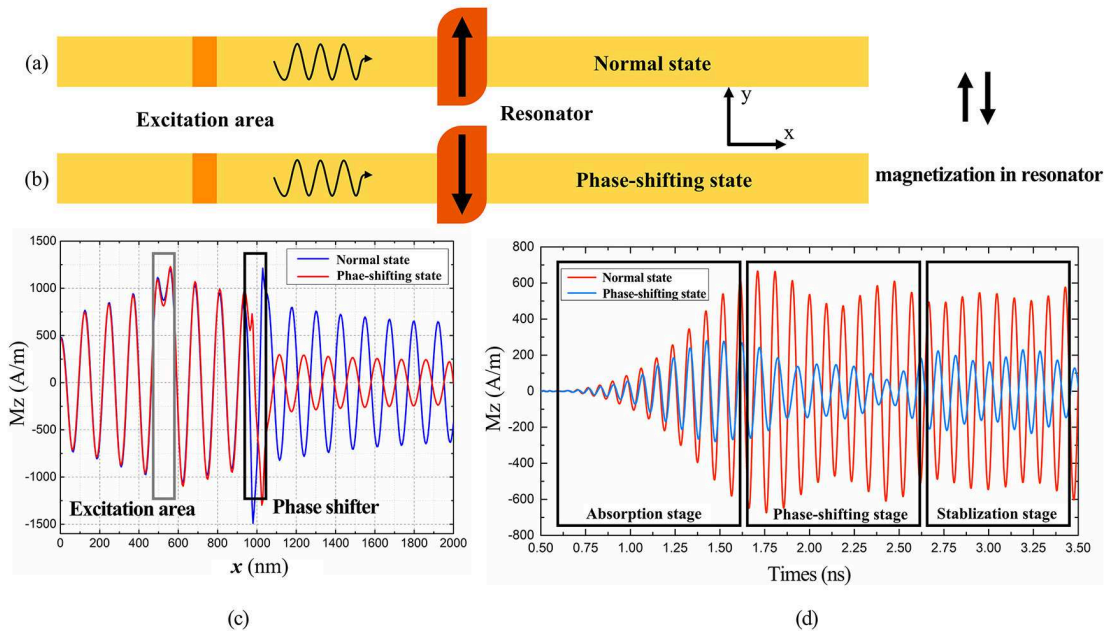}
\caption{\label{fig2} (a) The normal state with the magnetization of the resonator along +y direction. (b) The phase-shifting state with the magnetization of the resonator along -y direction. (c) The spatial distributions of z-component of the magnetization (Mz) in the waveguide along the x-axis under both states at 10 ns in the dynamic simulations. (d) The time dependent Mz at the point x = 1300 nm of the waveguide from 0.5 to 3.5 ns.}
\end{figure*}

\section{RESULTS and DISCUSSIONS}
\subsection{The phase-shifting mechanism}
The main part of the proposed magnonic phase shifter is the resonator. According to the magnetization status of the resonator, the beneath propagating spin waves in the waveguide have different temporal and spatial distribution. When the magnetization of the resonator points up as shown in Fig.2(a), the phase and amplitude of the SWs are rarely affected and shown by the blue line in Fig. 2(c). We name this state as normal state. However, when the magnetization of the resonator points down as shown in Fig.2(b), the SWs undergo a $\pi$-phase shift and an amplitude attenuation, as shown by the red line in Fig.2(c). We name this state as phase shifting state. To understand the phase-shifting mechanism clearly, Fig.2(d) shows the out-of-plane Mz component as a function of time at the point x = 1300 nm of the waveguide. We propose the phase-shifting process roughly has three stages. Firstly, resonator absorbs energy from the approaching SWs to excite oscillation (absorption stage). Secondly, when the absorption of SWs is strong enough, the resonance in the resonator occurs, and the phase of SWs start to vary (phase shifting stage). Finally, a steady state of resonance is achieved with a determined phase shift (stabilization stage).
\par
\begin{figure*}
\includegraphics{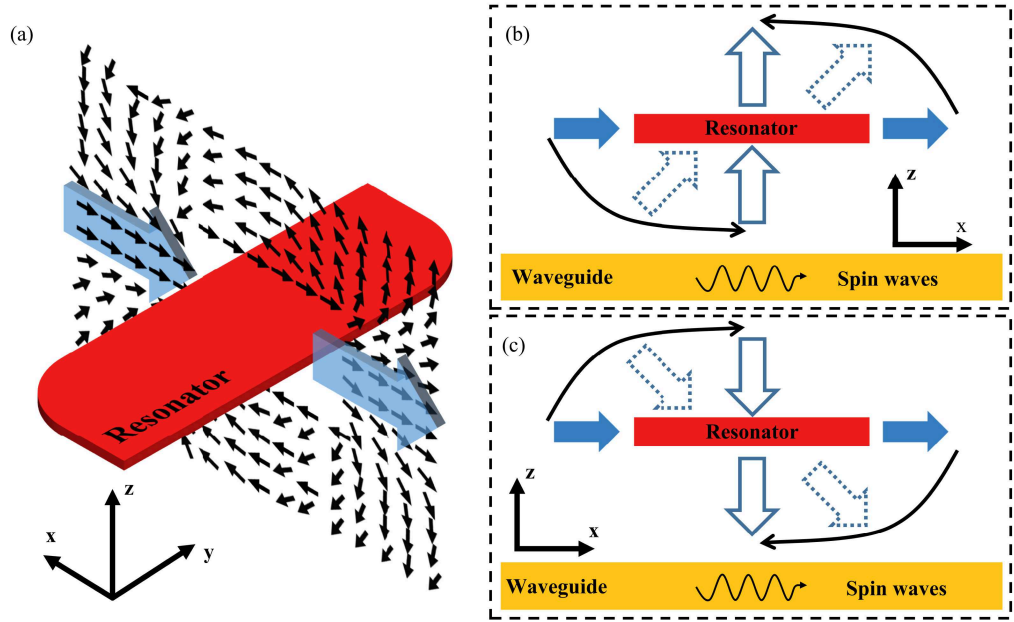}
\caption{\label{fig3} (a) Scheme of the dipolar stray field of the resonator. Simplified diagram representing the directions of the dynamic dipolar stray field and the SWs under (b) normal state and (c) phase shifting state.}
\end{figure*}

\begin{figure*}
\includegraphics{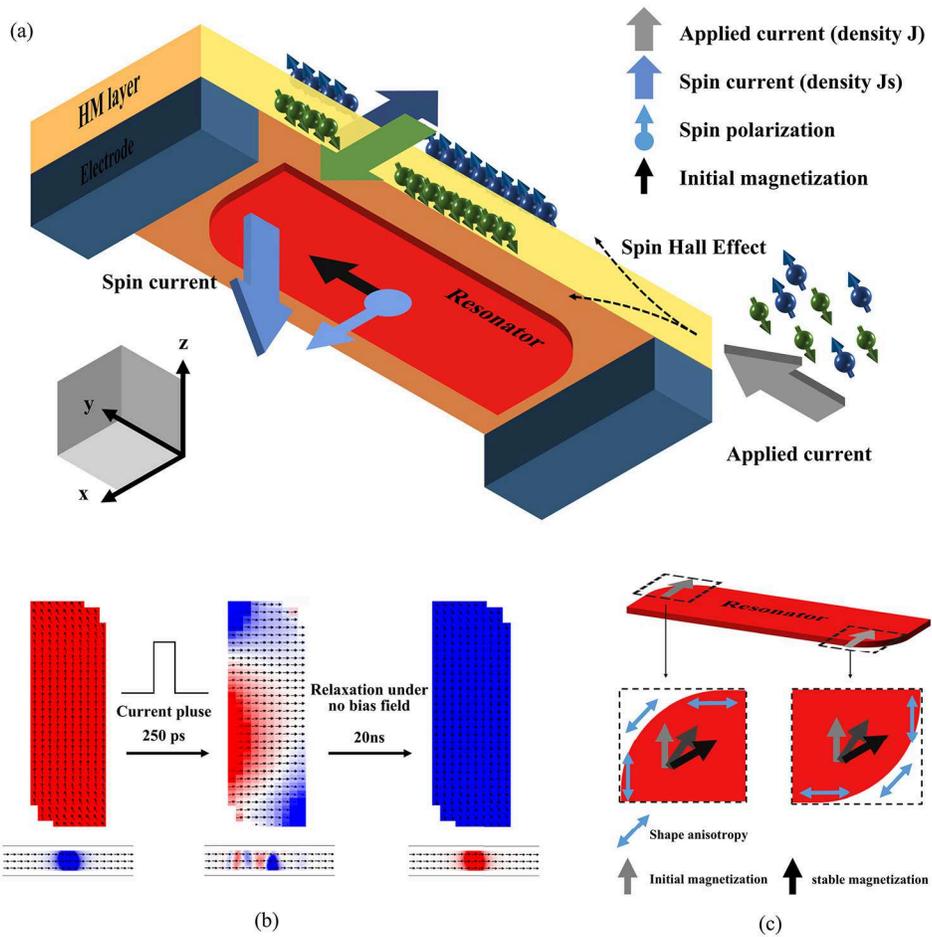}
\caption{\label{fig4} (a) Schematic diagram of the operational principle as a static phase shifter. (b) The switching procedure of the magnetization of resonator. (c) The shape anisotropy.}
\end{figure*}
The phase-shifting mechanism could be illustrated by a striking imbalance of dispersion relations of the SWs that propagate beneath the excited resonator \cite{47}. When the resonator is resonantly excited by the SWs, the dynamic dipolar stray field of the resonator, as shown in Fig.3(a), rotates around the surface. And through the dynamic dipolar stray field, the resonator exerts mutual influence on the incident signals. This interplay is affected by the chirality of the dipolar stray field. In the normal state (Fig.3b), the stray field on the waveguide is along the propagation direction of the SWs. While in the phase shifting state (Fig.3c), the stray field is opposite to the propagation direction of the SWs. Namely, the chirality of the stray field causes an imbalance of dispersion relations of the SWs. In other words, the propagation of SWs beneath the resonator is nonreciprocal, where the SWs could only be affect by the excited resonator in one magnetization direction. Remarkably, without varying the resonator, the original normal state as in Fig.2(a) will switch to phase-shifting state when the SWs reverse their propagation directions, and vice versa (see Appendix A). Hence, when the transmission direction of SWs is determined, we could then determine the states according to the magnetization of the resonator as discussed above.

\subsection{The application to a static phase shifter}
We first consider utilizing this magnonic phase shifter as a static controllable phase shifter. The operational principle is shown in Fig.4(a). To switch the magnetization of the resonator, we exert a current pulse into the heavy metal layer along the y-axis directions. When the current pulse flows into the heavy metal layer with strong spin-orbit coupling, a spin current is generated due to the spin Hall effect \cite{50,51,52,53}. The generated spin current is injected into the adjacent Py film in the x-axis directions. In the presence of the current-induced spin-orbit torque, the magnetization dynamics of the resonator is described by the LLG equation with an additional Slonczewski-like torque \cite{53}:
\begin{eqnarray}
\frac{{\partial m}}{{\partial t}} =  - {\gamma _0}m \times {H_{eff}} + \alpha m \times \frac{{\partial m}}{{\partial t}} + {\tau _{SLT}}
\end{eqnarray}
and
\begin{eqnarray}
{\tau _{SLT}} = {\gamma _0}{\tau _d}m \times (m \times \sigma )
\end{eqnarray}
where $m$ is the unit magnetization vector, $H_{eff}$ is the effective magnetic field, $\gamma_0$ is the gyromagnetic ratio, $\alpha$ is the Gilbert damping parameter, $\tau_{SLT}$ is the Slonczewski-like torque term, $\tau_d$ is the magnitude of $\tau_{SLT}$, and $\sigma$ is the direction of spin polarization. This torque term tends to align the magnetization towards the direction of spin polarization.
\par
It should be noted that the current density $J$ used in this paper is under the general simulation circumstance with the polarization degree P = 0.4 and the resistance mismatch $\Lambda$ = 2. These parameters give an ordinary polarization efficiency of the ferromagnetic films. However, the polarization efficiency could be much higher according to the spin Hall angle. The spin current density $J_s$ is related to the applied current density $J$ in spin-orbit coupling model as follow \cite{52}:
\begin{eqnarray}
{J_s} = {\theta _{SH}}(\sigma  \times J)	
\end{eqnarray}
where $\theta_{SH}$ is the spin Hall angle of the heavy metal layer, and the $\sigma$ is polarization of the electron spin. If the spin Hall angle of the chosen heavy metal material is large enough, the actually required current density $J$ can be very small.
\par
Here we give an example of the switching procedure based on micromagnetic simulations as shown in Fig.4(b). The left diagram represents the initial magnetizations of the resonator and the waveguide stabilized from ground state. The middle diagram is a snapshot of the magnetizations of the resonator and the waveguide after applying a current pulse with a density of 5 $\times$ 10$^{12}$ A/m$^2$ for 250 ps. Here the polarization direction is along the +x direction. After a sufficiently long relaxation (~20ns), the magnetization of the resonator is eventually toggled to the -y direction as shown in the right diagram. The relaxation process is affected by the shape anisotropy as shown in Fig.4(c), which ensure the certainty of the final magnetization status of the resonator. By reversing the current pulse direction, the magnetization switching can also be readily accomplished. The switching procedures are determinative stable. This operational mode is very suitable for the application of programmable circuits, since the device is reconfigurable and the energy consumption is extremely low.

\subsection{The application as a dynamic controlled phase shifter}

\begin{figure*}
\includegraphics{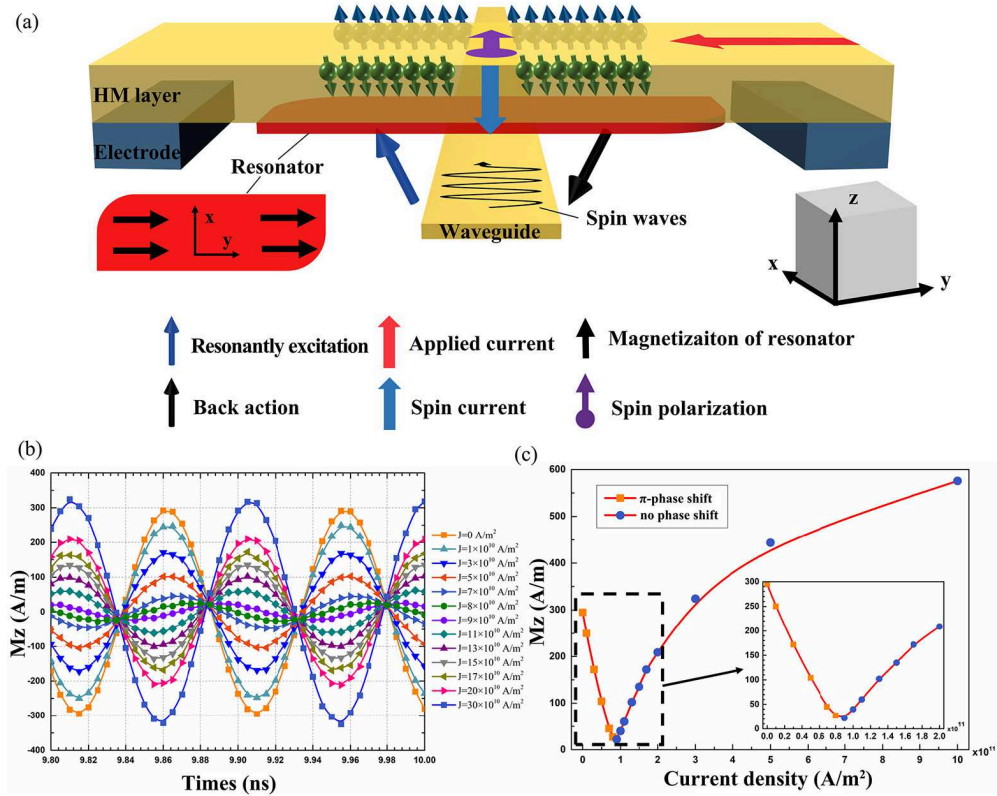}
\caption{\label{fig5} (a) Schematic diagram of the operational principle as a dynamic controlled phase shifter. (b) The graph shows the dependence of the waveforms of the SWs under different applied current densities. (c) The detailed relationship between the amplitude and the current density.}
\end{figure*}
\begin{figure*}
\includegraphics{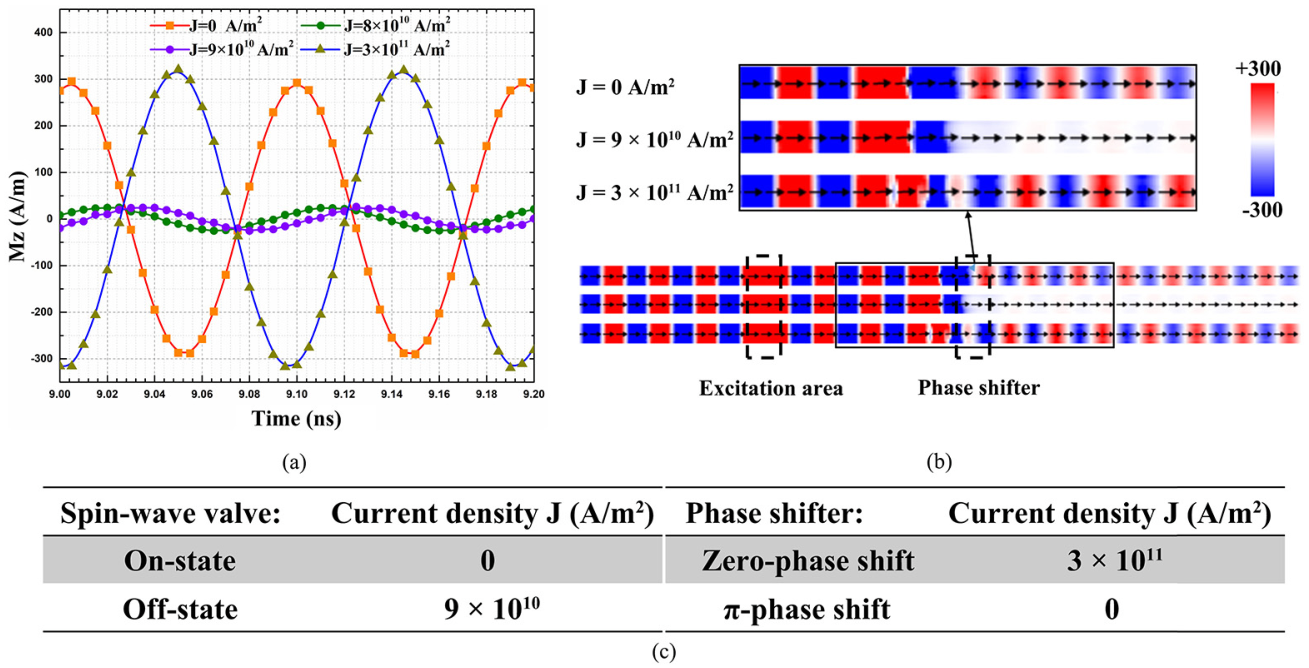}
\caption{\label{fig6} (a) The typical results of time dependent waveforms for applications of phase shifter or spin-wave valve. (b) The corresponding snapshots of the spatial distribution of dynamic magnetizations of the waveguide. (c) The table of the function of the device.}
\end{figure*}

The proposed magnonic phase shifter could also be utilized as a dynamic phase shifter in magnonic logic gate. The operational principle of the dynamic mode is depicted in Fig.5(a). In this mode, we maintain the magnetization of the resonator in phase-shifting state and exert constant currents (versus current pulse for static phase shifter) throughout the whole phase-shifting process. The polarization of generated spin current is along the +x direction. As mentioned above, such a spin current introduces a spin-transfer term to the LLG equation and thus alters the original excitation state. We observed a significant difference of the phase shifting effects at different current densities. Fig.5(b) represents the dependence of the waveforms of the SWs on different applied current densities J. The amplitudes of the SW signals differ remarkably, and there is a sudden change of the relative phase between $J$ = 8 $\times$ 10$^{10}$ A/m$^2$ and J = 9 $\times$ 10$^{10}$ A/m$^2$. The detailed relationship between the amplitude and the current density $J$ is summarized in Fig.5(c). The curve shows a single-valley image with a steep descending branch and a slowly ascending branch.
\par
We attribute this result to the interaction of the spin-orbit torque and the resonant excitation. In the initial stage, the spin-orbit torque tends to align the magnetic moments to the polarization direction. However, its strength is not enough to overcome the resonance process. The magnetic moments are easier to be resonantly excited even in presence of spin-orbit torques. As a result, the resonant excitation absorbs more energy from the SWs, and the amplitude of the SWs decreases obviously at the beginning, which is corresponding to the descending branch in Fig.5(c). As applied current density J increases, the procession of magnetic moments is gradually dominated by the spin-orbit torque which gradually diminish the resonant excitation in turn. Finally, the magnetic moments are pinned to the polarized direction by the current, and resonator has minute effect on the incident SWs as in the normal state. This process requires large applied current density, which corresponds to the slowly ascending branch. To confirm this mechanism, we did simulations by altering the spin current direction or magnetization of the resonator (see Appendix B).
\par
\begin{figure*}
\includegraphics{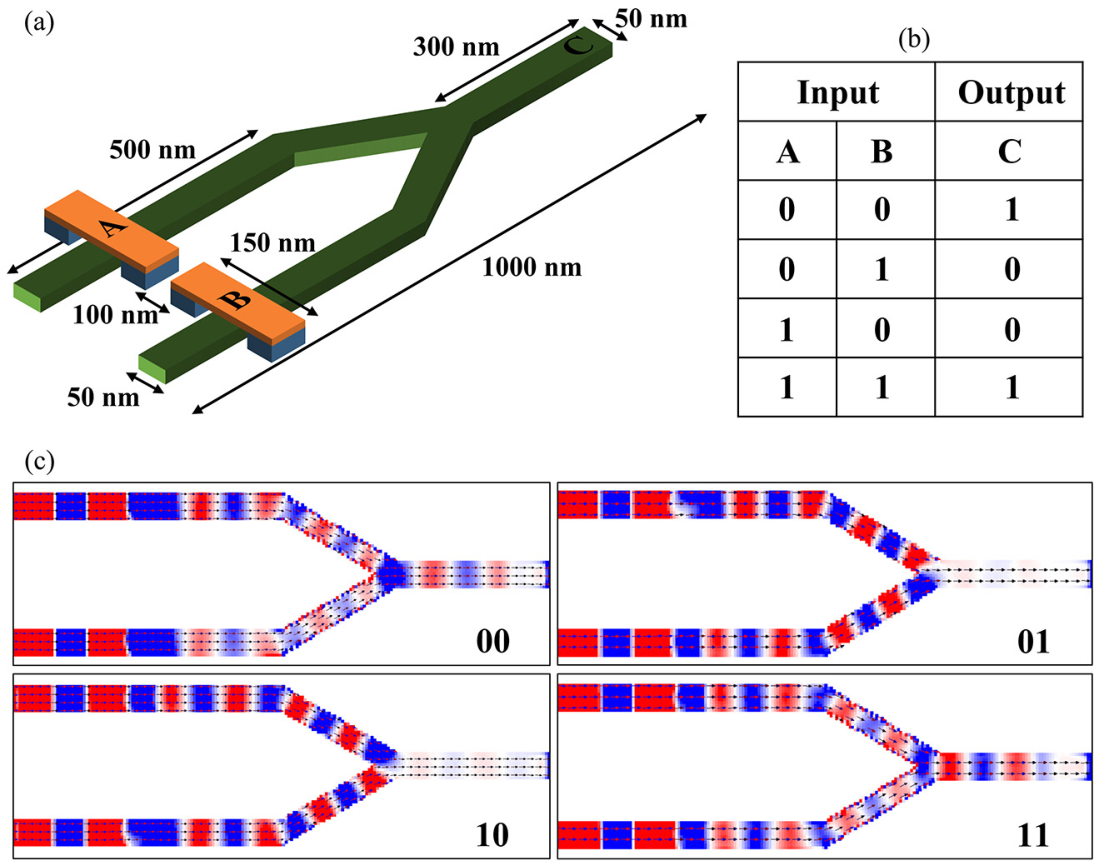}
\caption{\label{fig7} (a) Schematic diagram of the proposed XNOR gate. (b) The corresponding truth table of the device. (c) The simulated spatial maps of the logic gate under all logic inputs.}
\end{figure*}
\begin{figure*}
\includegraphics{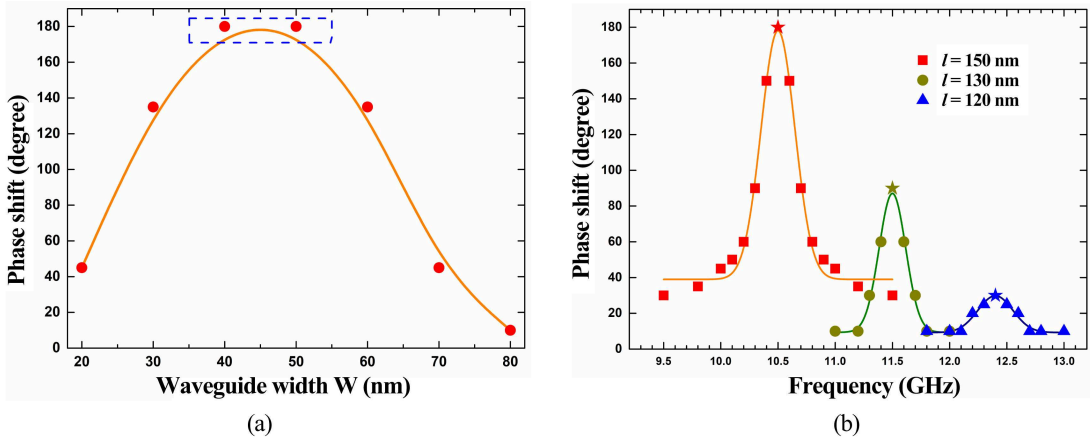}
\caption{\label{fig8} (a) The graph shows the change in SW phase shift with different waveguide width W. (b) The SW phase shifts are plotted as a function of the frequency under different resonator length $l$.}
\end{figure*}

\begin{figure}
\includegraphics{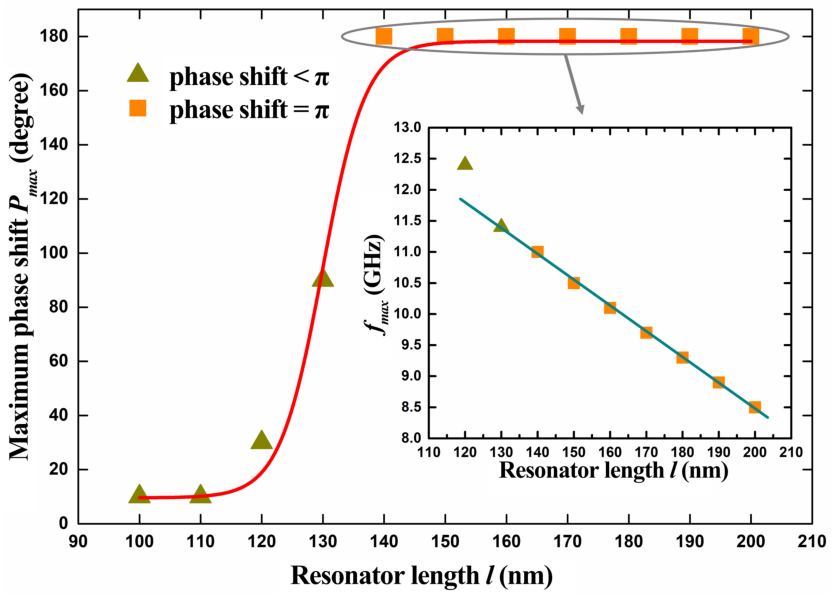}
\caption{\label{fig9} The dependence of $P_{max}$ on the resonator length $l$.}
\end{figure}
We could build a controllable spin-wave phase shifter or spin-wave valve based on this phenomenon. Fig.6(a) shows the typical results of time dependent waveforms. The corresponding snapshots of the spatial distribution of dynamic magnetizations of the waveguide are represented in Fig.6(b). At $J$ = 0 A/m$^2$, the incident SWs are shifted by $\pi$-phase when passing through the resonator. At $J$ =3 $\times$ 10$^{11}$ A/m$^2$, the phase of the incident SWs is unaffected with respect to the original signal. At $J$ = 9 $\times$ 10$^{10}$ A/m$^2$, the SWs are fully absorbed by the resonator that referes to an off-state. Hence, by only changing the applied current density, we could utilize this device as a dynamic phase shifter or spin-wave valve as illustrated in Fig.6(c).
\par
Accordingly, we demonstrate a XNOR gate using the magnonic phase shifter. Fig.7(a) represents the schematic diagram of the proposed XNOR gate. Fig.7(b) and (c) shows the truth table and the simulated spatial maps of the logic gate. The spacing between the two branches is wide enough to avoid dynamic magnetostatic coupling. SWs at frequency $f$ = 10.5 GHz are simultaneously excited in the left edge to ensure these two signals have the same initial phase. The resonators are the same as above mentioned to fit the application frequency. The input logic 0 and 1 represent the applied current density $J$ = 0 A/m$^2$ and $J$ = 3 $\times$ 10$^{11}$ A/m$^2$, respectively. The output logic 0 and 1 are defined as the destructive and constructive interference results, respectively. Both simulated snapshots and the truth table excellently demonstrate the operation of the XNOR logic.

\subsection{The application conditions}
To strengthen the applicability of the proposed magnonic phase shifter, it is necessary to investigate the application conditions of the device. To simplify the model and ensure the consistency of the arc edges of the resonator, we only consider the width of the waveguide W and the length of the resonator $l$ as the size variables. We firstly investigate the impact of waveguide width W on the phase shift under the original conditions. The result is shown in Fig.8(a). In a limited range (approximately from 35 nm to 55nm), the waveguide width W has minute effect on the phase shift. However, when W is either too narrow or too wide compared to the resonator length $l$, the phase shift amount decreases rapidly. Thus the SWs could hardly excite the resonator when W is too narrow. While the interplay from the resonator is not sufficiently strong to cause a $\pi$-phase shift when W is too wide. The results suggest that the ratio of W/$l$ should be chosen appropriately (about 1:3).
\par

For a given waveguide, it is critical to determine a suitable length of the resonator for the desired frequency. Fig.8(b) represents the frequency dependency of phase shift at different lengths $l$ (with W = 50 nm). For each curve, there exists a frequency marked by a star in the figure that have maximum phase shift. We define the frequency and the corresponding phase shift as maximum phase-shifting frequency $f_{max}$ and maximum phase shift $P_{max}$ respectively. The results show that only specific W and $l$ can achieve a $\pi$-phase shift for a frequency. Only when the $P_{max}$ is equal to $\pi$, the corresponding $f_max$ ($f_{\pi}$) is the exact application frequency. Fig.9 depicts the dependence of $P_{max}$ on the resonator length $l$. The inset represents the dependence of $f_{\pi}$ on $l$. Due to the limitation of dispersion relation of the waveguide (the cut-off frequency is ~ 8 GHz), $l$ $\textgreater$ 200 nm is not considered. The maximum phase shift $P_{max}$ decreases sharply near the point at x = 130 nm, due to the incompatibility of the resonator and the waveguide as also shown in Fig.8(b). In this model, the working frequency $f_{\pi}$ is linearly related to $l$ which could be summarized as follow:
\begin{eqnarray}
f_\pi = k*l + C
\end{eqnarray}

where $k$ = 0.04 GHz/nm and $C$ = 4.5 GHz according to the simulation. From the above formula, we could choose a suitable length of the resonator to fit the desired application frequency. The simulation results are in good agreement with the prediction equation (see Appendix C).

\section{CONCLUSIONS}
In conclusion, we propose a bias-free magnonic phase shifter based on a ferromagnetic resonator. This device has outstanding advantages in compatibility and miniaturization for its surface mounting structure. We use numerical micromagnetic simulations to show that the magnonic phase shifter could be utilized as a static element providing constant phase shift. Meanwhile, it could also be utilized as dynamic phase shifter controlled by the spin current. This phase shifter requires relatively low current input and provides stable phase shifts. We demonstrate a universal XNOR gate based on this phase shifter. Furthermore, the relationship between the application frequency and the size of the phase shifter is summarized. Results indicate that the magnonic phase shifter could be applied into a specific waveguide under desired frequency by selecting an appropriate size parameter. We believe this research could provide positive experiences for the development of the magnonic logic circuits.



%


\appendix
\section{The nonreciprocity of the resonator}
See Fig.10
\begin{figure}[H]
\includegraphics{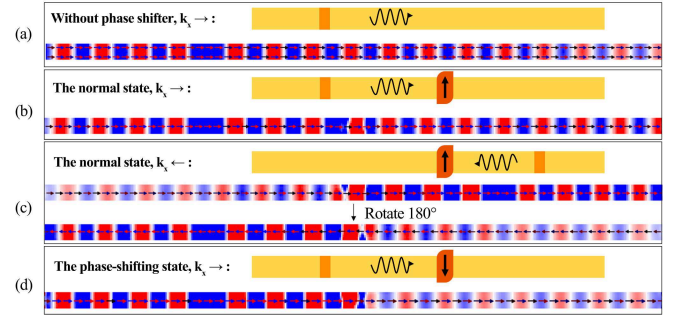}
\caption{\label{fig10} Demonstration of the nonreciprocity of the resonator. The snapshots of dynamic magnetizations of waveguide (a) without magnonic phase shifter at 5 ns, the direction of the wave vector $k_x$ points right; (b) with the phase shifter under normal state at 5 ns, the direction of the wave vector $k_x$ points right; (c) with the phase shifter under normal state at 5 ns, the direction of the wave vector $k_x$ points left; (d) with the phase shifter under phase-shifting state at 5 ns, the direction of the wave vector $k_x$ points right.}
\end{figure}
\section{Comparison of the phase-shifting effect under applying the spin current in different directions.}
In this section, we show that applying the spin current with spin polarization along –y direction could get the similar results under the phase shifting state, as shown in Fig.11. In this case, the density of the applied current needed to cut the SWs is less than the +x case. The obtained results conform to our expectation. The spin current in –y case also interacts with the resonant excitation and causes the similar affection. While the phase shifting results are unchanged if we perform this simulation under the normal state (not shown here). Under the normal state, the resonant excitation has little reaction to the incident SWs. Even if the input spin current affects the resonance process, there is no effect on the incident SWs.
\begin{figure}[H]
\includegraphics{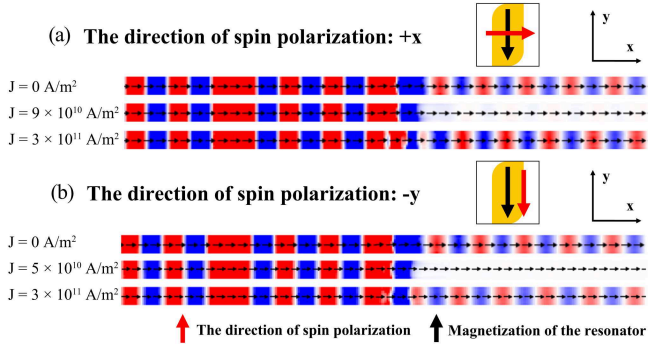}
\caption{\label{fig11} (a) The snapshots of magnetizations are shown under different current density with the spin polarization along +x direction. (b) The snapshots of magnetizations are shown with the spin polarization along -y direction.}
\end{figure}
\section{The detailed data of the phase shifting snapshots}
In this section, we give some simulated results mentioned in Fig.8 and Fig.9, see Fig.12 and Fig.13.
\begin{figure}[H]
\includegraphics{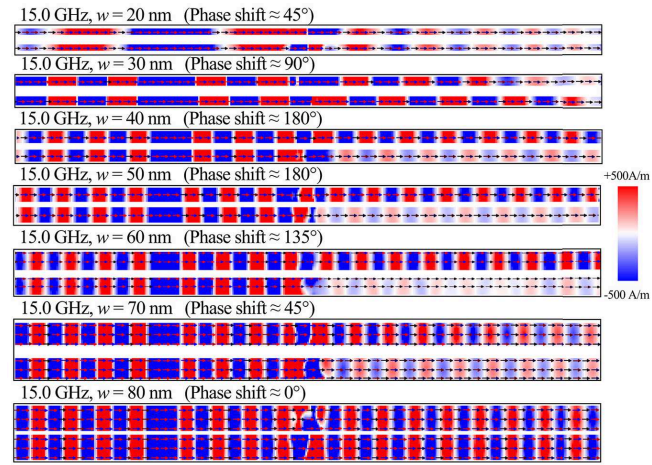}
\caption{\label{fig12} The snapshots of dynamic magnetizations represent the points in Fig.8.}
\end{figure}
\begin{figure}[H]
\includegraphics{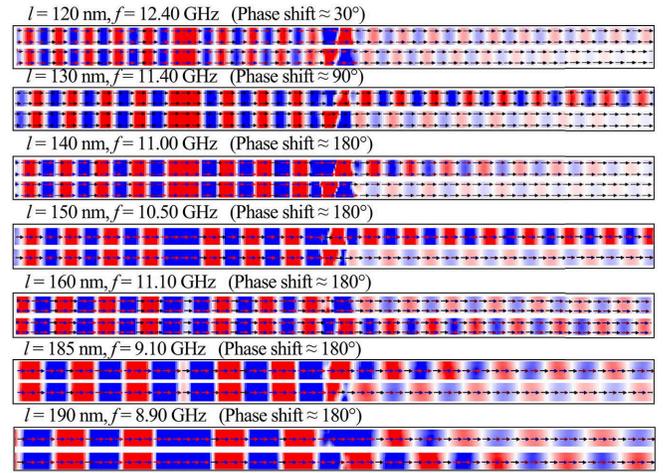}
\caption{\label{fig13} The snapshots of dynamic magnetizations represent the points in Fig.9.}
\end{figure}

\begin{acknowledgments}
This work is supported by the National Natural Science Foundation of China (grant Nos. 61734002, 61571079, 51702042 and 51827802), the National Key Research and Development Plan (No. 2016YFA0300801), and the Sichuan Science and Technology Support Project (No. 2017JY0002).
\end{acknowledgments}

\bibliography{zzk}

\end{document}